\documentclass[aps,prb,groupedaddress,showpacs,twocolumn]{revtex4-1}
\usepackage{graphicx}
\usepackage[
   colorlinks,        % Links ohne Umrandungen in zu wählender Farbe
   linkcolor=blue,   % Farbe interner Verweise
   filecolor=blue,   % Farbe externer Verweise
   citecolor=blue   % Farbe von Zitaten
]{hyperref}
\usepackage{textcomp}
\begin{document}

%\preprint{}

\title[Mn$^{2+}$/Mn$^{3+}$ state of La$_{0.7}$Ce$_{0.3}$MnO$_3$]
{Mn$^{2+}$/Mn$^{3+}$ state of La$_{0.7}$Ce$_{0.3}$MnO$_3$ 
by oxygen reduction and photodoping}


\author{Andreas Thiessen}
\author{Elke Beyreuther}
\email{elke.beyreuther@iapp.de}
\author{Stefan Grafstr\"om}
\author{Lukas M. Eng}
%\thanks{}
\affiliation{Institut f\"ur Angewandte Photophysik, 
Technische Universit\"at Dresden, D-01062 Dresden, Germany}

\author{Kathrin D\"orr}

\affiliation{IFW Dresden, Institute for Metallic Materials, Postfach 270116, 
D-01171 Dresden, Germany\\}

\affiliation{Institute for Physics, MLU Halle-Wittenberg, 
Von-Danckelmann-Platz 3, D-06120 Halle (S.), Germany}

\author{Robert Werner}
\author{Reinhold Kleiner}
\author{Dieter Koelle}

\affiliation{Physikalisches Institut and Center for Collective 
Quantum Phenomena in LISA$^+$,
Universit\"at T\"ubingen, Auf der Morgenstelle 14, D-72076 T\"ubingen, Germany}


\date{\today}

\begin{abstract}
Films of cerium-doped LaMnO$_3$, which has been intensively discussed as an 
electron-doped counterpart to hole-doped 
mixed-valence lanthanum manganites during the past decade, were analyzed 
by x-ray photoemission spectroscopy with respect to their 
manganese valence under 
photoexcitation.  The comparative analysis of the Mn 3s exchange 
splitting of La$_{0.7}$Ce$_{0.3}$MnO$_3$ (LCeMO) films in the dark 
and under illumination clearly shows that both oxygen reduction and illumination 
are able to decrease the Mn valence towards a mixed 2$+$/3$+$ state, independently 
of the film thickness and the degree of CeO$_2$ segregation. Charge injection from 
the photoconductive SrTiO$_3$ substrate into the Mn e$_g$ band with carrier lifetimes 
in the range of tens of seconds and intrinsic generation of electron-hole pairs 
within the films are discussed as two possible
sources of the Mn valence shift and the subsequent electron doping. 

\end{abstract}

\maketitle

\section{Introduction}
Mixed-valence manganites have been in the focus of extended research activities 
for decades due to their intriguingly manifold magnetic and electronic phases, 
high spin polarization, or the occurrence of the colossal magnetoresistance (CMR) 
effect in a number of such compounds \cite{ram97,coe99,hag03,dor06}. 
It is of fundamental academic interest to fully understand  microscopically 
the strong coupling of different degrees of freedom, 
the resulting subtle phase equilibria, the sensitivity 
of the latter to external 
stimuli, as well as other peculiarities such as electronic phase separation or 
completely new physical properties in thin film heterostructures. 
Furthermore, some compounds, such as 
La$_{0.7}$Sr$_{0.3}$MnO$_3$ (LSMO), show high spin polarization at room 
temperature, which makes them interesting for applications in 
upcoming oxide-electronic devices.

In the past, most frequently hole-doped rare-earth manganites such as 
La$_{1-x}$Ca$_{x}$MnO$_3$ (LCMO), the above-mentioned 
La$_{1-x}$Sr$_{x}$MnO$_3$ (LSMO), Pr$_{1-x}$Ca$_{x}$MnO$_3$ (PCMO), or 
Pr$_{1-x}$Sr$_{x}$MnO$_3$ (PSMO) -- all with 0$<$x$<$1, were investigated. 
There, in
the parent compounds (LaMnO$_3$ or PrMnO$_3$, respectively) part of 
the trivalent rare earth ions (La$^{3+}$ or Pr$^{3+}$) are 
replaced by divalent cations such as Sr$^{2+}$ or 
Ca$^{2+}$. To preserve charge neutrality part of the originally 
trivalent Mn ions are forced into a tetravalent 
state, finally leading to a mixed Mn$^{3+/4+}$ state, 
which corresponds to hole doping. 
The mixed Mn$^{3+/4+}$ valence
is crucial for the specific electronic transport mechanisms (e.g. the 
double-exchange scenario) of the manganites and the strong coupling 
of magnetic order and electric transport.

With regard to possible all-oxide or even all-manganite devices, the question 
whether LaMnO$_3$ accepts a La$^{3+}$ substitution by tetravalent 
cations such as Ce$^{4+}$, Te$^{4+}$, Sn$^{4+}$ was raised around a 
decade ago. Nominally such a substitution would result in a mixed 
Mn$^{2+/3+}$ state and electron doping. However, for the most intensively 
investigated case of Ce doping, it was found that single-phase materials 
can be grown only in the form of thin films in a limited range of 
doping concentrations \cite{ray03}, because of the unfavorable ionic sizes of 
Ce$^{4+}$ and Mn$^{2+}$. While the most previous investigations deal with 
fundamental preparation \cite{mit01c,yan04b}, transport \cite{man97,ray99}, magnetism 
\cite{ray99,geb99,lee01}, 
electronic structure \cite{min01,jol02,mit03a,han04a}, 
or valence \cite{phi99,kan01b} 
issues -- always motivated by the question whether an electron-doping is 
really present or not -- a few studies went a step further and focused on 
possible heterostructures including LCeMO films \cite{mit01a,mit03b,cho05}. 
A more detailed review 
of the debate was given in~\onlinecite{bey06}. 
In general, one can summarize, that La$_{1-x}$Ce$_{x}$MnO$_3$ (LCeMO) 
films tend to CeO$_2$ phase segregation and oxygen enrichment, the latter 
leading to an effective hole 
doping of as-grown films.

On the other hand, as discussed in our former XPS investigation \cite{bey06}, 
oxygen reduction by heating in an ultrahigh-vacuum environment is a 
suitable way to drive the Mn valence towards $2+/3+$, which is, however, 
connected with a decisive resistance increase and the loss of an important 
functional property, namely the 
manganite-typical metal-insulator transition (MIT) \cite{bey09}. 
As a further intriguing aspect, 
those oxygen-reduced, insulating, electron-doped LCeMO thin 
films exhibited a large photoconductivity and the recovery of the 
MIT, in contrast to the non-photosensitive as-grown hole-doped films.

Besides a number of microscopic explanations of intrinsic photoinduced effects 
in manganites (as summarized in \onlinecite{bey09,bey10}), the injection of 
photogenerated charge 
carriers from the substrate \cite{bey09,bey10,kat00a,kat01} seems to 
play a major and possibly 
technologically interesting role in thin-film-substrate heterostructures. 
In two studies of Katsu et al. 
\cite{kat00a,kat01} La$_{0.7}$Sr$_{0.3}$MnO$_3$ films on SrTiO$_3$ were 
illuminated 
with a broad-band white-light source and exhibited a negative photoresistivity (PR), 
according to the definition 
$PR=(R_{dark}-R_{illum})/R_{illum}$.

The negative PR was
interpreted as the injection of optically generated electrons from the 
SrTiO$_3$ substrate into the hole-doped film followed by the recombination 
of both carrier types in the film leading to a resistance increase. We 
picked up this stream of thinking in our two previous 
works, in which we observed a positive PR in 
La$_{0.7}$Ce$_{0.3}$MnO$_{3-\delta}$ \cite{bey09} and 
La$_{0.7}$Ca$_{0.3}$MnO$_{3-\delta}$ \cite{bey10}
films on SrTiO$_3$. By comparatively evaluating the wavelength dependence 
of the PR and the surface photovoltage we concluded that the 
photogenerated carriers must stem from interband transitions in the 
substrate or be excitated from interface states.

In the present work, we aim at deepening our understanding of the 
light-induced charge carrier generation 
in a tetravalent-doped thin-film manganite. We 
performed a comparative investigation of the Mn 
valence of LCeMO films of different thicknesses on SrTiO$_3$ under broadband 
white-light illumination, 
by evaluating the Mn 3s exchange splitting in the x-ray photoemission 
spectrum in order to clarify the 
doping type under photoexcitation. 

\section{Experimental}

\subsection{Samples}
\begin{table}
\caption{\label{tab_1}Samples of this study: 
LCeMO-film parameters; the full 
details to sample A can be found in \onlinecite{bey06}, the details 
concerning samples B, C, D in \onlinecite{wer09}. Note, that transmission 
electron microscopy had shown nanoscopic cerium oxide clusters 
in B and C, which were not visible in the XRD results.}
%\begin{indented}
%\lineup
\begin{ruledtabular}
\begin{tabular}{lrlcl}
label&thickness (nm)&p(O$_2$) (mbar)&XRD\\
\hline
A & 10 & 0.53 & single phase \\
B & 30 & 0.25 & single phase \\
C & 100 & 0.25 & single phase \\
D & 100 & 0.03 & CeO$_2$ cluster\\
\end{tabular}
\end{ruledtabular}
\end{table}

Four different LCeMO films, in the following labelled A, B, C, D, and summarized 
in table~\ref{tab_1}, were grown by pulsed laser deposition on SrTiO$_3$ 
(100) single crystal substrates. The substrates were 
10$\times$5$\times$0.5~mm$^3$ in size. 
Sample A was already the subject of our 
previous investigations -- the corresponding references \cite{bey06,bey09} 
contain the preparation details. 
The growth and structural analysis of samples B, C, D are given in 
\onlinecite{wer09}. Samples B and C were grown under identical oxygen 
partial pressure of p(O$_2$)$=$0.25~mbar but 
are different in thickness, while samples 
C and D have the same thickness but D was grown under a much lower 
oxygen pressure, which led to microscopic segregation of CeO$_2$ clusters. Thus, 
the comparison of B and C provides 
information on the thickness dependence of the 
observed effects, while contrasting C and D is relevant to quantify the 
influence of the phase 
segregation.

To separate the influence of oxygen stoichiometry from any of the 
above parameters, each sample was studied in two different states of 
oxygen reduction, which were prepared by heating in a low-pressure 
oxygen atmosphere: A \emph{slightly reduced} state was achieved by annealing the 
samples at 480$\symbol{23}$C in 10$^{-6}$~mbar oxygen partial pressure for 
1~h, while a \emph{highly reduced} state was prepared by heating 
at 700$\symbol{23}$C in 10$^{-8}$~mbar oxygen partial pressure for 2~h.

\subsection{XPS measurements}

The photoemission equipment for the XPS measurements uses 
a Mg anode and
has been specified elsewhere \cite{bey06}, where also details on the LCeMO
overview spectra, surface cleaning, and background correction are elaborated. 
All spectra were recorded at room temperature.

For studying the effect of photoexcitation on the Mn core signals the 
respective sample was simultaneously 
illuminated by the broadband 
output of a Hg arc lamp providing a maximum intensity of 18.6~\textmu W~mm$^{-2}$ 
(integrated over the whole spectrum). Between 390~nm and 250~nm the 
integrated intensity measures 9.6~\textmu W~mm$^{-2}$.
This interval is the spectral range relevant for band-to-band excitation of 
electrons in the SrTiO$_3$ substrate.
The white light was focused by a fused-silica lens and transmitted through 
an UV-transparent viewport into the UHV apparatus. 

In contrast to our former XPS work \cite{bey06}, which was limited to one 
10-nm-thick, gradually oxygen-reduced LCeMO film, now the Mn 3s exchange splitting 
energy $\Delta E_{3s}$ is measured for a whole set of LCeMO films, 
cf.~table~\ref{tab_1}, in two different states of oxygen reduction.

As pointed out earlier, the Mn valence $V_{Mn}$ can be calculated from the 
measured value of $\Delta E_{3s}$ via the empirical equation:

\begin{equation}
\label{eq_valence}
V_{Mn}=9.67 - 1.27 \mbox{(eV)}^{-1} \cdot \Delta E_{3s} \quad ,
\end{equation}

which is based on XPS results of a number of Mn compounds with well-known Mn valences
\cite{zha84,gal02}.

\begin{figure}
\centering
\includegraphics[width=0.45\textwidth]{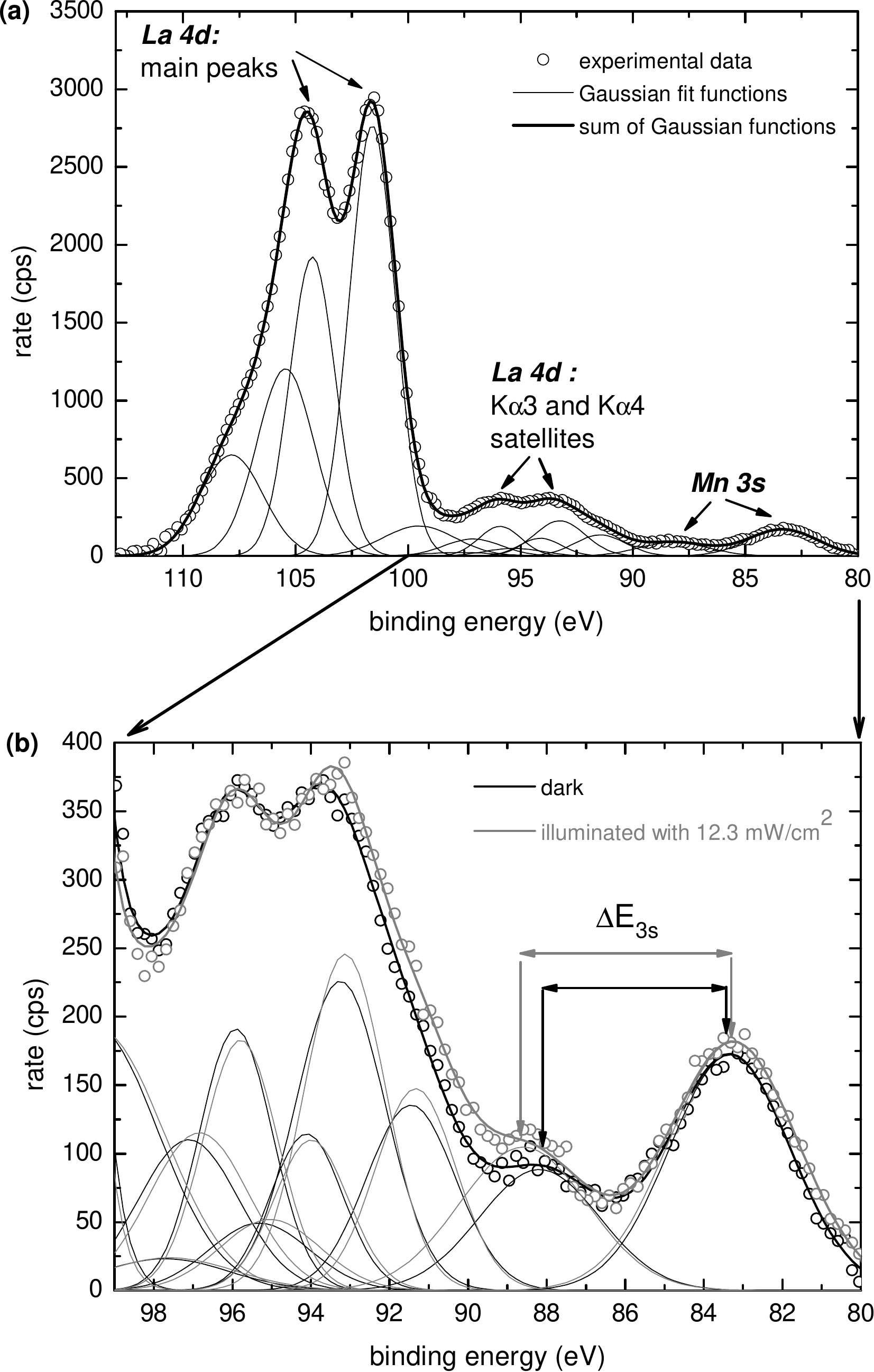}
\caption{\label{fig1}(a) Mn 3s doublet and neighboring La 4d signals: 
Deconvolution into 
individual peaks in order to find a realistic value of the Mn 3s exchange splitting energy.
(b) Increase of the Mn 3s exchange splitting energy 
under illumination in an oxygen-reduced LCeMO film.}
\end{figure}

As shown in figure~\ref{fig1}(a), the intensity of the Mn 3s lines is comparably 
low and, more problematic, they are located in the direct neighborhood of the 
La 4d lines and part of their satellites. For large Mn 3s exchange splittings 
the smaller peak of the Mn 3s doublet even sits on the shoulder of the 
La 4d K$\alpha_3$ satellite. Consequently, the quantitative determination of 
$\Delta E_{3s}$ and thus $V_{Mn}$ needs a careful deconvolution of the spectrum 
into the individual peaks of the La 4d and Mn 3s structures. The La 4d structure 
consists of four peaks and the corresponding K$\alpha_3$ and K$\alpha_4$ 
satellites. Together with the two peaks of the Mn 3s level this gives 14 peaks, 
each described by three parameters (position, height, and width). Thus, 
we obtain a regression 
function with 42 variables, which are fortunately not completely
independent. 

The La 4d signal is made up of  
four peaks with a fixed energetic separation and an intensity ratio between 
La 4d$_{5/2}$ and La 4d$_{3/2}$ of 0.69, which results in eight independent 
fit parameters, see~\cite{how78}. Since the K$\alpha_3$ and K$\alpha_4$ satellites are 
energetically shifted images of the original La 4d signal, they theoretically cannot 
have free parameters. However, due to the noise of the measured data, it 
was necessary to treat the intensity and width of La 4d$_{5/2}$ K$\alpha_3$
and the width of La 4d$_{3/2}$ K$\alpha_3$ as free parameters initially -- in 
order to make the regression algorithm more flexible. In 
practice, first the regression analysis \cite{fityk} (based on the 
Levenberg-Marquardt algorithm) 
was performed for the whole La 4d structure, then the regression function 
was extended by the two Mn 3s peaks and the regression analysis was 
rerun. Figure~\ref{fig1}(a) represents a typical result of the fitting procedure.

After the respective oxygen reduction 
procedure the sample was moved 
from the preparation chamber to the photoemission chamber -- all in the 
same UHV apparatus-- and an overview x-ray photoemission 
spectrum as well as a detailed spectrum of the region of 
a possible C 1s peak (as a test for possible organic contaminations) were taken. 
To study the influence of illumination on the Mn valence and to conclude on 
the nature of the photogenerated carriers,
spectra of the La4d/\-Mn~3s region in the dark and afterwards under 
illumination were recorded. Figure~\ref{fig1}(b) depicts two exemplary spectra 
of this region with and without light; the increase of the exchange splitting, 
which corresponds to a decrease of the Mn valence towards 2$+$/3$+$ 
according to (\ref{eq_valence}), is clearly visible.

\begin{figure}
\centering
\includegraphics[width=0.45\textwidth]{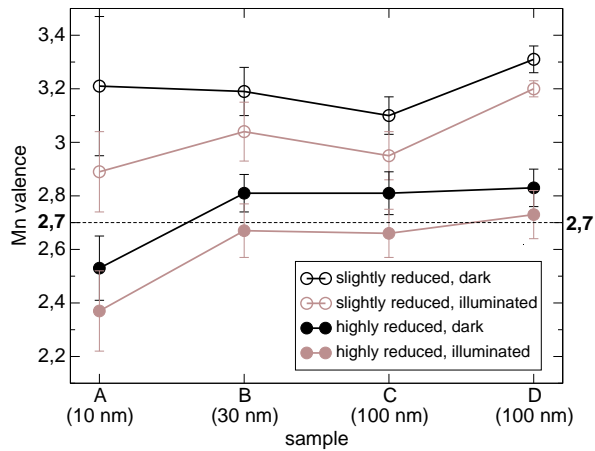}
\caption{\label{fig2}Overview of the main results: Mn valence of the four LCeMO films 
in two states of oxygen reduction, in the dark and under illumination, respectively. 
For better orientation, the nominal value of 2.7 is depicted as dashed line.}
\end{figure}

Due to the weakness of the Mn 3s signal, noise influences the 
fitting results. Depending on the concrete case, each measurement was repeated 
between 5 and 13 times. 
For each of those measurements the sample was freshly prepared. Note that 
heating in oxygen-rich atmosphere approximately recovers the as-prepared state, 
as shown formerly \cite{bey06}. In 
some selected cases, a second dark measurement was performed after 
the measurement under light excitation. There was no remarkable
difference observed between the two dark values of the valence shift. Thus we can 
exclude that the valence shifts are caused by further outdiffusion of 
oxygen into the UHV surroundings. Finally, the mean values of 
$V_{Mn}^{dark}$, $V_{Mn}^{illuminated}$, and 
$\Delta V_{Mn}=V_{Mn}^{dark}-V_{Mn}^{illuminated}$ were statistically
calculated based on 95\% confidence intervals. Figure~\ref{fig2} 
and table~\ref{tab_2} show the mean values 
and the confidential intervals (errors).

In the dark, as clearly visible in figure~\ref{fig2}, 
the slight reduction procedure is not 
able to drive the Mn valence below $+3$, while the stronger reduction 
leads to Mn valences below $+3$ and thus indeed to an electron doping. This is 
in qualitative accordance with our previous study \cite{bey06}. Within the 
error bars, it is hard to see systematic differences between the four individual 
samples. Besides the fact that the lowest Mn valence is achieved in the thinnest 
film, no serious systematic dependence on the thickness or the degree of 
phase segragation is visible.

Under illumination, both the slightly and the highly reduced films exhibit a \emph{further} 
decrease of the Mn valence towards the mixed 2$+$/3$+$ state, which is equivalent 
to a photoinduced increase of the electron density in the Mn-$e_g$ orbital.

\begin{table}
\caption{\label{tab_2} XPS results of the light-induced Mn-valence shift 
$\Delta V_{Mn}$, 
the amount of the corresponding injected charge carriers $\Delta N_{e_g}$ according 
to eq.~(\ref{eq_N_eg}) and the associated carrier lifetime $\tau$. The 
latter are calculated 
using eq.~(\ref{eq_time}) and assuming carrier injection from the substrate 
to be the dominant mechanism.}
%\begin{indented}
%\lineup
\begin{ruledtabular}
\begin{tabular}{clccc}
 sample  & reduction & $\Delta V_{Mn}$ &  $\Delta N_{e_g}$ ($\times$10$^{15}$) 
 & $\tau$ (s)  \\ 
\hline
A  		& slight 	& -0.32 $\pm$ 0.09 & 2.8 $\pm$ 0.8 & 2.9 $\pm$ 0.8 \\
(10 nm) & high 		& -0.16 $\pm$ 0.08 & 1.4 $\pm$ 0.7 & 1.5 $\pm$ 0.7 \\
\hline
B  		& slight 	& -0.15 $\pm$ 0.11 & 3.9 $\pm$ 2.9 & 4.1 $\pm$ 3.0 \\
(30 nm) & high 		& -0.14 $\pm$ 0.04 & 3.6 $\pm$ 1.1 & 3.8 $\pm$ 1.2 \\
\hline
C  		& slight 	& -0.15 $\pm$ 0.05 & 13.8 $\pm$ 4.2 & 13.8 $\pm$ 4.2 \\
(100 nm) & high 	& -0.15 $\pm$ 0.08 & 13.8 $\pm$ 7.4 & 13.8 $\pm$ 7.4 \\
\hline
D  		& slight 	& -0.11 $\pm$ 0.06 & 9.5 $\pm$ 5.3 & 9.5 $\pm$ 5.3 \\
(100 nm) & high 	& -0.11 $\pm$ 0.03 & 9.5 $\pm$ 2.5 & 9.5 $\pm$ 2.5 \\
\end{tabular}
\end{ruledtabular}
\end{table}

Importantly, control experiments with only-cleaned films 
in the dark and under 
illumination showed a mixed 3$+$/4$+$ valence \emph{and no change} under illumination.

\section{Discussion}

In general, at least two scenarios have to be discussed as possible explanations 
for the observed Mn valence decrease under illumination. First, the photoinduced 
generation of electron-hole pairs within the LCeMO films and a delayed recombination 
due to hole-trapping at oxygen vacancies, as already observed for other 
manganite compounds \cite{gil00}, would be conceivable.
Second, as pointed out earlier, for the 10-nm-thick LCeMO film, carrier injection 
from the substrate was suggested to be the main origin for the 
previously observed photoconductivity \cite{bey09}. 
Here, all films show a similar shift of the Mn valence 
towards the electron-doped state under illumination. Thus it seems
likely that the injection mechanism plays a role also for the 
thicker films. Assuming a semiconductor band model, which has been successfully 
employed to explain a number of photoexcitation effects in manganites in the 
past, the samples are 
heterostructures of a wide-gap semiconductor (SrTiO$_3$, E$_g$=3.2~eV) and 
a narrow-gap 
semiconductor (LCeMO, E$_g$ around 1.0~eV), showing a band bending at 
the interface leading to a 
built-in field. Electron-hole pairs which are generated by photoexcitation 
in the substrate are separated by this field and the electrons can be 
injected into the LCeMO film.

In the following, the results for $V_{Mn}^{dark}$, $V_{Mn}^{illuminated}$, 
and $\Delta V_{Mn}$ are used to calculate the carrier densities 
$n_{e_g}$ in the Mn-$e_g$ orbitals of the LCeMO films according to:

\begin{equation}
\label{eq_n_eg}
	n_{e_g}=\frac{3-V_{Mn}}{a^3}  \quad .
\end{equation}

The lattice constant $a$ measure 0.388~nm for samples A--C and 0.389~nm for 
sample D \cite{wer09}. The total number of charge carriers injected from the 
substrate into the film, $\Delta N_{e_g}$, can be calculated as:

\begin{equation}
\label{eq_N_eg}
	\Delta N_{e_g}=\frac{\Delta V_{Mn}}{a^3}\cdot V  \quad ,
\end{equation}

with $V$ being the volume of the film. The results for 
$\Delta N_{e_g}$ are listed in table~\ref{tab_2}. Note that
equations~(\ref{eq_n_eg}) and (\ref{eq_N_eg}) are valid for the 
charge-injection as well as for the intrinsic-carrier-generation scenario. 

Assuming $\Delta N_{e_g}$ to be the equilibrium number of 
photogenerated electrons in the film, the absorbed photon 
flux $\Phi$ and the life time 
$\tau$ are connected with $\Delta N_{e_g}$ via:

\begin{equation}
\label{eq_time}
	\Delta N_{e_g}=\Phi \cdot \tau  \quad .
\end{equation}

Using the charge-injection scenario, we can calculate $\tau$ from 
this relation. However, we have to be aware of the fact that the formula 
gives only a rough estimate of the carrier life 
times, since eq.~(\ref{eq_time}) is only valid for a quantum efficiency 
for the electron-hole 
pair generation in the substrate of 100\% and a simple exponential relaxation 
process. However, with the known flux of photons with energies above 
the SrTiO$_3$ bandgap of $\Phi=$9.45$\times$10$^{14}$~s$^{-1}$ and under the 
assumption that all photons from the light source are absorbed in the 
SrTiO$_3$ bulk, 
life times between 1 and 10~s are calculated, see table~\ref{tab_2}. Obviously, 
the life time increases with the film thickness. Within the mechanism postulated 
above this can be understood as follows: the spatial separation of the electron-hole 
pairs by the internal field prevents the electrons from recombining. 
For thicker films the mean distance of the electrons from the interface is 
larger and 
consequently the life time is longer. Similarly long life times (22~s)
were also observed by Gao et al. in electron-doped La$_{0.8}$Te$_{0.2}$MnO$_3$ 
films on SrTiO$_3$ by analyzing the relaxation of the photoresistivity \cite{gao09}.
For the intrinsic-carrier-generation scenario, a serious value for $\Phi$ 
cannot easily be determined, since the LCeMO absorption coefficients for the 
differently oxygen-reduced states are unknown. Thus, we are reluctant to 
estimate life times for this scenario.

Furthermore, we have to comment on why the illumination-induced 
Mn valence shift is exclusively seen in oxygen-reduced samples. In principle, 
this observation is compatible with both scenarios. For the intrinsic 
carrier generation scenario, the
carrier (hole) trapping at oxygen vacancies is indeed essential.
If there were no centers 
for pinning the holes, a fast recombination of the photogenerated electron-hole pairs 
could take place and would prevent the films from any electron doping -- 
exactly as observed in the as-prepared case.
However, the charge-injection scenario would be possible in 
the as-prepared (oxygen-rich) samples as well. At the current state of knowledge 
one may only speculate that the probably dramatically changed band 
alignment at the interface -- note that SrTiO$_3$ 
changes from p-type in the ideal stoichiometric case to n-type under reduction -- 
leads to very different recombination conditions. 

Finally, possible systematic errors of the Mn valences 
due to the surface sensitivity of XPS 
measurements have to be considered. Taking into account that only the 
uppermost 2--3~nm of the films are probed and that former XAS results \cite{wer09}
revealed a higher Mn valence deeper inside the films than at their surfaces, the 
mean Mn valence can be higher than the values of table~\ref{tab_2}. The 
finding that $\Delta V_{Mn}$ is independent of the film thickness 
might be, at least partially, an artifact produced by this systematic error.

\section{Summary and Outlook}

The Mn valence of 10-, 30-, and 100-nm-thick La$_{0.7}$Ce$_{0.3}$MnO$_3$ films 
was determined from the Mn 3s exchange splitting in the 
x-ray photoemission spectrum in the dark and under white-light illumination. 
Independent of the film thickness and the degree of CeO$_2$ segregation, both 
oxygen reduction and illumination turned out to be effective ways to drive 
the Mn valence towards 2$+$/3$+$ and thus make the LCeMO films electron-doped. 
Nevertheless, the two routes are not equivalent: While \emph{oxygen reduction} 
alone drives the system towards a Mn valence below $+$3, \emph{photoexcitation} 
lowers the Mn valence only in films with a certain \emph{initial} degree of 
oxygen reduction. 

Discussing (i) a scenario postulating charge injection from the photoconductive 
SrTiO$_3$ substrate into the Mn e$_g$ band and (ii) a scenario assuming intrinsic 
photostimulated carrier generation and subsequent hole-trapping at oxygen 
vacancies within the LCeMO films as possible (and maybe coexisting) 
origins for the decreased 
Mn valence in LCeMO, we estimated the number of photogenerated injected carriers and 
their lifetimes in the films.
To completely clarify which of the scenarios is the dominating one, 
a similar investigation of LCeMO on a 
non-photoconductive substrate as well as a more extended study of the 
spectral dependence of the photoconductivity than in ref.~\onlinecite{bey09} 
would be illuminative.

The question whether electron-doped manganites can be established by 
tetravalent-ion doping can be answered as follows: It is not 
primarily the tetravalent doping ions but post-deposition oxygen reduction 
and -- optionally -- illumination which lead to an electron-doped state. 
The fact that reduction and illumination can induce an electron-doped 
state in divalent-ion-doped manganites (whose growth is commonly easier) 
as well \cite{bey10} questions the need of tetravalent-ion substitution 
in manganites.

\begin{acknowledgments}
This work was financially supported by the German Research 
Foundation (DFG, project no. BE 3804/2-1 and EN 434/31-1).
\end{acknowledgments}

%\bibliography{LCeMO_revisited}
%\bibliographystyle{apsrev4-1}


\end{document}